# AN OVERVIEW OF PHISHING VICTIMIZATION: HUMAN FACTORS, TRAINING AND THE ROLE OF EMOTIONS

Mousa Jari[1, 2]

[1]School of Computing, Newcastle University, Newcastle, UK
[2]College of Applied Computer Science,
King Saud University, Riyadh, Saudi Arabia

***ABSTRACT***

*Phishing is a form of cybercrime and a threat that allows criminals ('phishers') to deceive end-users in order to steal their confidential and sensitive information. Attackers usually attempt to manipulate the psychology and emotions of victims. The increasing threat of phishing has made its study worthwhile and much research has been conducted into the issue. This paper explores the emotional factors that have been reported in previous studies to be significant in phishing victimization. In addition, we compare what security organizations and researchers have highlighted in terms of phishing types and categories as well as training in tackling the problem, in a literature review which takes into account all major credible and published sources.*

***KEYWORDS***

*Phishing, emotion, information, victimization, training.*

## 1. INTRODUCTION

Phishing is a kind of social engineering attack that is used to steal an individual's data, including personal identification details, credit card numbers or any other credentials. This activity occurs when a phisher pretends to be someone who is a trusted individual and persuades a victim to open a certain email or a message. When a victim opens such a communication, his information can be hacked/leaked and made available to the email/message sender. McAlanay and Hills described phishing as a social engineering tool or a threat that causes a risk to cyber security [3]. They further highlighted that phishing emails or messages are based on the assertion of some urgency or threat where an attacker or phisher causes a victim to become blackmailed having been encouraged to respond to the email or message accordingly [3]. According to Shaikh et al., phishing is a serious threat in the cyber world that is causing billions of dollars of losses to internet users through the use of social engineering and technology by gaining access to their financial information [4]. In phishing, the attacker sends spoof emails to the internet user which deceives victims and causes them to disclose their sensitive and confidential data [4]. Consequently, from an analysis of the relevant research based on the common elements which have been identified, one can define phishing as a cyber threat which, due to the deployment of social engineering techniques and technological means, leads to messages and emails being sent to internet users resulting in the retrieval of personal information about victims, hence causing them monetary or other damage through the leaking of information.

In this research, the aim is to identify human factors, and specifically emotional variables, which lead to a higher probability of phishing victimization. This problem has been discussed in various studies, and so the method employed in this paper is to analyse the literature review and secondary research in order to highlight the emotional factors which play a significant role in phishing victimization, as well as comparing how security organizations define and address phishing and provide advice on how to avoid becoming a victim.

## 2. OVERVIEW OF PHISHING

Phishing is a relatively new concept which was first employed in the late 1990s, and in the past years there has an increasing trend of damage caused by phishing. Rather than simply the deployment of technical expertise to attempt to successfully compromise system security, phishing can also be defined as a semantic attack that uses social engineering tactics to persuade internet users to disclose their private and confidential information such as login credentials, social security numbers, and bank account details. Phishers most commonly use an e-mail which includes an embedded hyperlink with a message either sharing some threat, such as a warning message about account closure, or reporting positive news, for example hinting at an unclaimed reward, to attract a potential victim. When a person clicks on the malicious link, it leads to a web-based form that mimics those of valid and authentic websites asking a user to enter login credentials. Once added, such information is then used to compromise network security and thus sensitive information reaches the phisher. When phishers have retrieved sensitive information from victims, they can then either sell the information, open bank accounts, or even steal the victim's money. Such phishing attempts are believed to be the 'vector of choice' among cybercriminals.

The Anti-Phishing Workgroup has discovered up to 40,000 phishing websites per month, targeting almost about 500 unique brands; however, the US Department of Defense and the Pentagon have reported more than 10 million phishing attacks each day, which is clearly a huge number [4][6]. However, Harrison also highlighted the fact that the success rate of phishing attacks is never 100% and often varies between 30-60% [4][6]. This research study aims to explore what makes these attempts successful, focusing on the emotional factors that may lead users to share confidential information with someone unknown to them.

### 2.1. Types of Phishing According to Research and Security Organizations

Researchers and the security sector have listed various types of phishing through which the phishers target internet users. In a paper entitled 'Fifteen years of phishing: can technology save us?' Furnell et al. highlighted four major types: spear phishing, clone phishing, whaling, and bulk phishing [3][5]. In spear-phishing, specific individuals or companies are targeted using a tailored message. In this type of scam, the attacker is more likely to have some background information about the target, based on which the message that is created becomes more convincing and successful in deceiving the recipient. Therefore, users become more likely to be targeted and lose their confidential information. Meanwhile, clone-phishing attackers make use of a valid email that contains a URL or attachment and retain the content of the actual message. However, the embedded link or attachment is replaced with a malicious file so that the original sender is spoofed due, for example, to a claim that the email message is an update of an earlier version. In contrast, whaling is considered to be a particularly threatening type of phishing where CEOs or other senior or high-value individuals in an organization are targeted. Here, the medium used in communicating the message is still the email, but in addition similar kinds of threats are also sent through 'vishing' (voice phishing) and 'smishing' (SMS phishing) which are terms used to specify threats via voice telephony and text messaging. The final category of bulk phishing occurs when there is no specific target or any tailored message. The approach employed instead is to send bulk emails to as many users as possible, and the success of this kind of scam depends on such large-scale mailing where a sufficiently large number of recipients mistakenly believe that the email is relevant to them [3][5].

Various security organizations have categorized phishing in different ways, but the forms listed overlap with the four major types indicated above. Reports published by the US Federal Trade Commission, the Surveillance Self Defense and Get Safe Online groups, and Phishing.org specify 14 major types of phishing, including: spear phishing, session hijacking, email spam, content injection, web-based delivery, phishing through a search engine, link manipulation, vishing, smishing, key logging, malware, trojans, ransomware, and malvertising [5–10]. The common element in all phishing categorizations provided by security organizations and Furnell et al.'s research is the fact that attackers use email, voicemail, or SMS to accomplish phishing [5]. Similarly, victims in all types vary from ordinary internet users to specific companies or high-profile individuals. In addition, the concepts exploited in all the phishing types overlaps with the categories recognized by the security organizations, except for web-based delivery, phishing through a search engine and key logging. In web-based delivery phishing, which is also called 'man-in-the-middle' phishing, the attacker is positioned between the customer and the original website. Using his phishing system or network, the phisher identifies the confidential information of the victim during the completion of a deal between the user and the legitimate website. In the case of search engine-based phishing, the user is captured by being (re-)directed to websites purportedly offering low-priced products or services. When the user tries to purchase a product by adding his credit card details, the data are collected by the phisher. Moreover, key logging phishers identify keyboard strikes and mouse clicks performed by a user, and from this information they manage to retrieve passwords and other confidential data[3][5].

## 2.2. Training and Increasing Awareness

Online safety and the avoidance of all threats that remain present around users is not easy. In the case of phishing, this is a specific kind of scam that plays with human psychology and attracts the attention of victims using various techniques to cause the damage explained above. Questions therefore arise concerning how people can be trained to stay safe from phishing, and publications from various researchers and security organizations addressing this issue are compared here in order to draw conclusions about how to protect users against the rising threat of phishing.

Jensen et al. considered aspects of training which might mitigate the impact of phishing attacks, focusing on the use of 'mindfulness' techniques [9][11]. The authors specified that simple decision making or mental shortcut methods to avoid phishing are no longer effective, since not only are they short-term strategies but also phishers are now very familiar with such models. So, the researchers wanted to create an innovative approach to training which teaches internet users to develop new mental models and strategies for the allocation of attention when examining online messages. Rather than a rule-based approach which repeats multiple rules and cues, the training was designed as an exercise to enhance the degree to which users attend to and understand the approach being used in the received message; in other words, to promote 'mindfulness' in evaluating the message. The concept of mindfulness here concerns paying receptive attention to one's experience and surroundings so as to improve the ability to understand one's environment along with an enhanced self-regulation capacity and stronger behavioural control. Their training module consisted of graphics in addition to text promoting mindfulness, and helped provide a better understanding of how to avoid phishing attacks, since the graphical representation of concepts is thought to enhance the capacity to acquire information leading to better performance in complicated tasks [9][11]. The project aimed to provide participants with a blend of mindfulness training techniques and a rule-based approach so that they could respond to phishing attacks more effectively. To test the effectiveness of the training, a dummy phishing attack was launched in which the participants were directed towards a fictitious website where they were asked to enter their university account login credentials. The results indicated that the graphic and text-based training formats was equally successful in decreasing participants' probability of responding to phishing messages in such a way as to become victims. However, the approach using mindfulness significantly reduced the likelihood of participants responding to the phishing messages, and hence was found to be useful against phishing [11].

Wash and Cooper have also explored the training models that can work best against phishing [12]. The researchers indicate that raising awareness among users through facts-and-advice training or storytelling models works better in combating phishing, but using professional security experts or peers to deliver such training is needed to make it more effective. The methodology of the study involved sending 2000 participants phishing emails to gather data necessary to assess the effectiveness of the activity. The lessons that could be learned included to "type in URLs; don't click on them" and "look for HTTPS", that "misspellings can signal fake emails" and "phishing is your problem; don't rely on others to protect you". In total, 17 lessons were compiled from the results of the activities performed. It was found that not all of the lessons helped in combatting phishing attacks, but all were considered significant enough to be presented to the participants when using the fact-telling approach. Meanwhile, the comparison with a story-based strategy led to the surprising outcome that the facts-and-advice approach resulted in fewer clicks leading to scams when an expert was used for the training, whereas the storytelling approach also resulted in lower click rates but only when peers were used rather than experts [12]. From the above-mentioned studies it can be concluded that multiple factors may lead a user to become caught by a phishing attack, and that appropriate training and education is needed in order to be safer.

With respect to the benefits of such education, the conclusions drawn in a study by Chaudhary include a number of significant recommendations as follows [13]:

I. Providing any new knowledge in order to be up to date is important, but security education should also result in eliminating misconceptions relating to security.
II. The security education that should be part of a curriculum needs to be up-to-date, and it should cover both new technologies, and information about sophisticated phishing threats and attacks.
III. Security education must impart knowledge related to technological and non-technological attacks and threats.
IV. The design of curricula for security education should be based on the input of relevant stakeholders, including teachers, learners, and IT and security professionals, since their experience, skills and knowledge can help in covering a wide range of security-related topics.
V. The adoption of a more interactive way of teaching and learning methods can be quite helpful in making both security learning and teaching more interesting and potentially effective [13].

In parallel with academic researchers, many security organizations have also put a lot of emphasis on user security against phishing attacks, because its severity can vary from password retrieval to stealing money, ultimately causing considerable damage. Among the most prominent security organizations is the National Cyber Security Centre (NCSC), which is a UK-based organization that provides support to critical organizations, including many in the public sector and industry as well as the general public.

In response to the rapid increases in cyber threat levels, the NCSC provides efficient incident responses to mitigate harm and facilitate recovery, and compiles information on the lessons learned that can be useful in the future. Apart from providing solutions to possible phishing threats, NSCS is also concerned with educating people to develop self-reliance against phishing attacks. For this purpose, a major contribution of the Centre is the design of practical resources for school students who take an interest in cyber security studies. The projects on which NCSC is working to provide cyber security education include the CyberFirst courses, schools, colleges, bursaries and apprenticeships and associated resources, and the CyberSprinters programme [14].

Similarly, the Get Safe Online organization provides users with a greater awareness of phishing and online scams through its informative online articles and shares tips and tricks to raise consciousness among users concerning phishing attacks. Advice is given, for example, on how to use emails wisely, how to identify fraudulent emails, how to distinguish between legitimate and phishing websites and emails and, if one has lost money due to an online scam, what course of action to take [10].

The Surveillance Self-Defense organization is also based on providing general public protection from phishing attacks. Its literature specifies the intensity of malware and its role in introducing phishing threats. The implementation of malware by phishers is usually based on stealing passwords, where the malware is installed when a user opens or clicks on a malicious link, downloads an unknown file, visits a compromised websites, downloads automatic content, or even when USBs are shared while plugging into suspicious ports. However, despite the multiple ways through which malware can be used for phishing [9], users can be educated to avoid being a phishing victim by implementing five important measures:

1. Updating systems and using licensed software.
2. Backing up data.
3. Pausing before clicking, and thus to be more vigilant and to avoid clicking immediately.
4. Using full-disk encryption along with a strong password.
5. Using better anti-virus techniques.

From the discussion above and in the light of the relevant research, it can be concluded that the frequency and intensity of online scam, phishing, and fraud activity are increasing with the passage of time, and so, in order to be safe, security education and training are necessary and perhaps should be mandatory.

### 2.3. The Role of Emotion

Chaudhary has emphasized the role of emotion in his research [13], explaining that the manipulation of emotion is generally found to be a prime target of phishers. Ignorance, a desire to be liked, gullibility, and wanting to be helpful to others are among the aspects associated with emotionality which are commonly targeted by scammers or phishers, who rely on the exploitation of vulnerability and weakness. People are found to be more inclined towards sharing their information with others when strong emotions have been triggered, and human behaviour when triggered this way is more likely to be driven by subconscious processes. The problem here is that the functionality of the subconscious mind is not based on logical or analytical behaviour, a fact which is exploited by phishers in pursuing their aims [13]. However, although emotions are very important and can be used against a victim as a weakness, they can also act in the victim's favour as a strength too. If the emotions which are exploited by phishers instead remain under the control of the user, this may help to combat phishing, which implies that emotions should also have a significant role in training and awareness-raising.

Chaudhary has discussed a very interesting type of phishing and social engineering attack in his research. This is called farming, where a phisher develops a relationship with a victim and continues to obtain relevant information over a certain period of time [15][13]. This activity is usually conducted in four phases. The first phase is information gathering, which involves the collection of the necessary data so that a relationship with a potential victim can be built. The second phase is based on developing the relationship, such as by coordinating with the victim and building a trust-based connection. In the third phase, exploitation starts where the victim is manipulated and deceived to obtain critical desired information, and the final phase is the execution of an attack using the information provided, to the detriment of the victim but beneficial to the attacker.

Emotion may exert a significant influence on many human cognitive processes such as attention, perception, memory, learning, reasoning and questioning, and problem-solving. If a person can manage to understand his emotions and learns how to control them, he can understand his surroundings better, communicate more efficiently, and even appreciate the worth of any relationship [16]. In relation to phishing, it can therefore be proposed that, if internet users are provided with the training and awareness based on emotional control, then they will be less likely to be successfully targeted by phishers.

To mitigate the malevolent exploitation of emotions, Jaeger and Eckhardt have highlighted the significance of emotions in awareness-raising and training [15]. They believe that human emotions are learnt, and when they are taken under appropriate control the impact of phishing attacks may be overcome. The researchers analyzed the relationships among the constructs of protection-motivation theory (PMT) Nomology [22] that involve fear and the motivation for protection and in actual security-related behaviour, indicating that perceived threat perceived coping efficacy in response to threat encourages a person's motivation towards self-protection in combatting the threat. So, when an individual encounters a phishing attack and faces its likely consequences; then, after being threatened, he starts to believe that he can respond to the situation using learnt behaviours and emotions which can ensure protection against such threats in the future [15]. Moreover, this helps not only in terms of training but also in creating awareness among peers. Such learning can also lead to technical solutions, such as users protecting themselves from phishing by implementing technical countermeasures including deciding not to click on any unknown or potentially malicious link, not downloading an .exe file, and deleting any dubious email or sending it to the junk folder.

## 2.4. Human Factors in Phishing Victimization

If one asks what makes phishing successful or what causes victims to become entrapped in phishing, the answer is simple: the victim himself. More specifically, it can be said that phishers target the victim's emotions which they manipulate to achieve their aims [16]. The above sections have indicated that emotions have a major role in phishing attempts, and the focus of the remaining discussion is to answer the research question of the study: what are the human factors, and specifically the emotional variables, which lead to phishing victimization.

In considering the nature of emotions and other psychological variables, Chaudhary cited various aspects of the human psyche which play a major role in phishing victimization [13] and specified several psychological states and factors that are mainly targeted by phishers which may lead a user to comply with the instructions given as part of the phishing attempt [13]. These include:

i) Reciprocation: where potential victims are more likely to comply with malicious instructions when they have a feeling of gratitude towards the phisher and feel that they are granting a favour to one in need.

ii) Consistency and commitment: since people like to be seen as trustworthy by fulfilling promises. If this trait is targeted by the phisher, to make one feel that he has made a promise, then it is possible that the person may comply with the phisher's instructions and demands.
iii) Social proof: people may be deceived more easily if they are provided with persuasive evidence, such as being convinced that one is not alone in doing something and everyone else is doing the same thing, so that trapping a victim becomes more likely.
iv) Liking: using the emotion of liking someone is often exploited as a tool by phishers, because people more readily comply with someone they like. If a phisher manages to masquerade as a person the victim likes, the phishing attempt could succeed.
v) Authority: people generally comply with authority, since being a responsible citizen usually means complying with an authorized person. So, if a phisher manages to appear authoritative, he can use the victim's tendency to comply with the demands of an authority to manipulate him.
vi) Scarcity: if a phisher manages to convince his target that something he wants is in short supply and will not be available afterwards, then it is more likely that the victim may comply with the phisher's instructions.

In addition to the above-mentioned psychological states and emotions described by Chaudhary, Vishwanath et al. considered the dimensions of the email and social media behaviour of individuals which result in getting trapped by phishing attempts [17]. They highlighted the fact that social media users, and especially those who regularly check Facebook notifications, are more likely to be targets of social media phishing. However, social media are quite distinct in providing relational information which can help in the detection of deception. Social media-based phishing attacks are multi-staged in the sense that the user receives a friend-request followed by messages. This is in contrast to email-based phishing which is single-staged, where a phisher uses a persuasive subject line which either causes a feeling of fear in cases of a threat, or a sense of happiness following a piece of fake news such as concerning winning a lottery or an amount of money. Vishwanath et al. concluded that users with low levels of emotional stability are more likely to start worrying and lose their emotional control based on the subject of the email. This kind of behaviour can create impulsive email habits. For example, in response to the sound of a single email notification, a user may react by checking and immediately opening an email to answer it, which may lead to reactively clicking on malicious phishing links [17]. Responding to emails with feelings of being nervous, curious, happy or under threat has also been explored by other researchers because this area of research has shown considerable promise.

Abroshan et al. recently conducted a noteworthy research study regarding human behaviour and emotions which influence the success of phishing attacks' [18]. They highlighted previous studies which have found that emotional behaviour can significantly affect responses to phishing emails, and proceeded to develop a holistic method including the use of psychological and phishing mitigation to identify highly susceptible users in organizations who are at the risk of clicking on phishing emails. Their proposed solution is comprised of three modules involving behaviour measurement, risk scoring and mitigation, and the system can be delivered online. It is also a flexible solution which can be expanded by adding more human factor root-causes; for example, "more behavioural and emotional factors that might impact falling into a phishing scam" (p. 349). This study significantly highlights the importance of human behaviour and emotions in relation to security behaviour such as the propensity to get caught up in a phishing scam. For example, the emotions of users such as fear and anxiety, especially in certain situations like the Covid-19 pandemic, can play a pivotal role in making phishing attacks successful. This is because the user's awareness and knowledge of security can be overshadowed due to the emotion of fear [18]. In reacting, users might click on a suspicious phishing link without thinking, supposing that the information is required due to Covid-19 health impacts.

## 2.5. User Knowledge, Education, and Understanding

Dealing with phishing attempts can be controlled through the use of software; however, the best prevention can only be provided through the user's improved knowledge, education and understanding. In one study, Arachchilage and Love emphasized that anti-phishing education and knowledge needs to be considered in order to combat phishing [19], and they investigated the extent to which procedural knowledge or conceptual knowledge has positive effect on users' self-efficacy to be safe from phishing attacks. Using a theoretical model based on Technology Threat Avoidance Theory, data was collected from 161 computer users who were provided with a questionnaire to get their feedback. It was found that both procedural and conceptual knowledge positively impacted the users' self-efficacy, ultimately resulting in the enhancement of their phishing threat avoidance behaviour. A later study by He and Zhang supported the claim that users' knowledge, education and understanding play a significant role in repulsing phishing attacks, and the authors concluded that "Training programs and educational materials need to relate cyber awareness to employees' personal life, family, and home, in order to be more engaging and to encourage employees to change their cybersecurity behaviour" [20].

Subsequent research regarding knowledge capabilities was conducted by Wash et al., who surveyed 297 participants with matching demographic characteristics in the US, allowing them to share their experiences of phishing emails [21]. This study provides evidence that humans may perceive and experience phishing emails in a very idiosyncratic manner, using different capabilities and knowledge in contrast to technical filters. For example, their past knowledge may assist them in detecting and becoming suspicious of phishing attacks, such as their familiarity with previously received emails as well as their expectations regarding incoming emails. It can be assumed that this knowledge is contextual and every individual will have a unique set of relevant experience, which will be utilised to detect missing and unexpected informational units in emails. Because technical solutions seldom spot these types of phishing attacks, such knowledge-based information residing only in the human mind is critical in spotting phishing attacks, whereas technical expert-based filters lack this information processing capability. For example, humans can use their knowledge to conduct an investigation or delay the response to an email and request further information from the sender [21]. This shows that the user's knowledge is critically important in combatting phishing attacks.

From a linguistic perspective, although it is acknowledged that those who are non-native speakers are more vulnerable to phishing attacks [23], most published studies fail to consider language-based phishing vulnerability. However, Hasegawa et al. conducted a noteworthy online survey of 302 Japanese, 276 South Koreans and 284 German participants representing a total of 862 non-native English speakers. The results of the analysis of data revealed that participants who were not familiar or confident with the English language had a high propensity to ignore all emails written in the English language [23]. Additionally, a qualitative analysis revealed five key factors that aroused the concern of participants in identifying phishing emails in English. These include difficulties in identifying errors in the language, unfamiliarity with the written English in phishing content, difficulty in understanding English content, and decreased attention. These findings suggest that it would be necessary to develop different strategies to tackle the susceptibility to phishing emails among non-native speakers, as well as to consider the importance of language barriers when formulating interventions to assist non-native speakers to combat phishing attacks.

## 2.6. Demographics Factors

In addition to the emotional factors discussed by Chaudhary and others, various demographic variables are believed to be significant in phishing victimization. However, other demographic characteristics have been found to have an impact on resilience against phishing. For example, Gopavaram et al. found that phishing resilience and age have a negative relationship, so that older users are more likely to become confused about the legitimacy of genuine and phishing websites. However, no significant relationship was identified between phishing resilience and gender [24], whereas Sheng et al. [25] found that age and gender are key demographic factors that can indicate the levels of susceptibility to phishing. Their analysis indicated that women click on malicious links provided in phishing emails more often than men, and hence are more likely to provide phishers with confidential information. Such differences in gender-based behaviour might be based on the role of technical education, since males often have more technical knowledge than females. Age was also found to have a significant relationship with phishing susceptibility, and participants between the ages of 18-25 years were more likely to fall into phishing traps. But this factor was also linked with levels of education, since participants in the age group concerned were found to have received a relatively lower level of education, limited training material, fewer years spent using the Internet, and low capacity in risks management [25]. So, in relating demographics to phishing, Arachchilage and Love's conclusion that improved education can reduce phishing victimization was supported.

### 2.7. Online Habits and Behaviour, Responding Impulsively to Emails, and the Role of Mental Models

In research where the scope and purpose are to understand the phishing attacks and victims' reasons for opening malicious links, an investigation of the mental models used and online behaviour exhibited is very significant. A simple explanation of a mental model is an individual's own thought processes in relation to how a particular phenomenon works in a real-life scenario. Mental models are based on an individual's learning, experience, skills and knowledge which improve the thought process concerned, hence resulting in some specific behaviour or outcome. So, no single mental model can work against phishing but several mental models can serve the purpose, and Jaeger and Eckhard have explained that schemata and mental models are key elements used to achieve high levels of awareness [15]. Critical cues that are needed to activate both of those mechanisms should be based not only on the characteristics of emails but should also be linked to security warning alerts. Meanwhile, mental models specifically relating to phishing could be more complex, since they may vary depending on the type of phishing attacks concerned. An individual's mental model is shaped by past experience where phishing plays a major role in obtaining situational information relating to security awareness. It was concluded that experienced users signify their level of awareness by using the security-related information cues available, which shows that the experience helps in developing schemata and remembering the critical cues, ultimately leading to pattern matching and improved thought processes in working against phishing [15]. In another study, Sibrian et al. conceptualized the thought processes and human behaviour involved using a model of the social decision-making process divided into two systems [26]. The first was defined as the source of emotional reactions based on experience, which works quite rapidly and almost impulsively with very little voluntary control, whereas the second system is based on reasoning, focus, and choice. The operations linked to this second system require attention, and can be disrupted when it is reallocated or disturbed. Since it is more logical and rational, whereas the emotional system is more impulsive, phishers exploit it so that the other system either does not react or takes too much time to respond. This is how phishers exploit the human mental model, due to which an individual may feel fear, happiness, curiosity or even urgency to respond to the phishing message quickly, which affects the overall user's behaviour while responding to a phishing attack and leads to getting entrapped in a phishing attack [26].

### 2.8. Expert versus Non-expert Thought and the Ability to Detect and Avoid Falling Victim to Phishing

The behaviour of experts and non-experts to phishing attacks differ, which primarily depends on one's skills, knowledge and experience. Nthala and Wash explain that non-experts follow four sense-making processes according to which they determine if the email they receive is in actuality, a phishing message [27]. In the first stage, they identify, without going into detail, whether or not the email is relevant to them. In the second stage, the goal is to understand why they received the email. Here, non-experts try to understand the email in more depth. In the third stage, they start to take a positive action using a sense-making process to fulfil the request made in the email. In the last stage, either the email is closed, deleted, or even moved to re-reading. The core element of this stage is the sense of marking the task as completed by closing the email [27].

As opposed to non-expert behaviour, Wash highlighted that, to identify a phishing email, experts follow a three-stage process [28]. In the first stage, experts tend to consider why such an email has been received and how it is relevant to them, and identify any discrepancies. In stage two, they may entertain suspicions about the email by analysing its features such as the presence of a link requiring a click. Here, they manage to identify that the email is based on a phishing message. In the third step after this thought process using their mental model, experts deal with the phishing email either by reporting or deleting it [28].

## 3. CONCLUSIONS

Phishing is one of the most common problems which causes an individual to lose confidential information such passwords or credit card numbers which may even trigger the theft of money. In this paper, we compared what security organizations and researchers have emphasised in terms of phishing types and categories as well as training and awareness in tackling the problem. Numerous emotional factors are targeted by phishers; however, with the help of training and anti-phishing education, emotions can be managed and controlled, and self-control can be developed that can lead to phishing attempts being unsuccessful.